# Does one still need to "shut up and calculate"?

Masud Mansuripur
James C. Wyant College of Optical Sciences, The University of Arizona, Tucson

**Abstract**. In learning quantum mechanics, an essential question has always been: How does one go about developing a "physical feel" for quantum phenomena? Naturally, one needs a basis or ground zero to start from, and that basis must be unlike anything with which we are already familiar in consequence of our experiences with the world of classical physics. We argue (channeling Richard Feynman) that the most elementary and the least cumbersome concept to build upon is the existence of complex probability amplitudes for physical events. An event that can take place in multiple alternative ways should be treated by adding the corresponding amplitudes when the paths are, in principle, indistinguishable, and by adding the probabilities themselves when the paths are distinguishable. Once we accept this principle and hone our intuition by examining quantum phenomena in its light, we will be on the path to "understanding" quantum mechanics. Elementary examples from the field of quantum optics demonstrate how adherence to Feynman's principle could lead to a better, more "intuitive" appreciation for the magic of quantum mechanics.

**1. Introduction**. Several years ago, the late Professor Willis Lamb (Physics Nobel Laureate, 1955) gave a seminar entitled "*What if Isaac Newton had discovered Schrödinger's equation?*" Lamb's contention was that the passage of time would have enabled us by now to develop an intuitive feel for quantum mechanics, in the same way that Newtonian mechanics — along with the rest of classical physics — makes intuitive sense to us today. The essential question, of course, has always been: How does one go about developing a "physical feel" for quantum phenomena? Naturally, one would need a basis, or ground zero, or square one, to start from, and that basis must be unlike anything with which we are already familiar in consequence of our experiences with classical physics and with other physical sciences.

In this paper, we argue (channeling Richard Feynman) that the most elementary and the least cumbersome concept to build upon is the existence of complex probability amplitudes for physical events.[1-3] The fact that an event that can take place in multiple alternative ways should be treated by adding the corresponding amplitudes when the paths are, in principle, indistinguishable, and by adding the probabilities themselves when the paths are distinguishable, is that holy grail. Once we accept this principle and hone our intuition by examining quantum phenomena accordingly, we will be on the path to fulfilling Willis Lamb's premise. The following sections present a few examples from the field of quantum optics[4-8] to demonstrate how the application of Feynman's principle could lead to a better, more "intuitive" appreciation for the magic of quantum mechanics.

Section 2 provides an elementary treatment of the problem of photon redistribution at a regular beam-splitter (RBS). We begin by assuming that two nearly identical single-mode wavepackets, one containing two photons in the number-state $|2\rangle$, the other, a single photon in the number-state $|1\rangle$, simultaneously arrive at the entrance ports of an RBS. We proceed to invoke Feynman's principle to derive the probability amplitudes for the various photon-number-states that emerge at the splitter's exit ports. The important notion of Bose enhancement emerges naturally from this analysis.[1] In the end, we confirm that the same result can be obtained by an application of the standard algebra of photon creation and annihilation operators,[6-8] which, although powerful and fairly straightforward, lacks the transparency and the intuitive appeal of the more elementary method of calculation. (Both types of analysis can be readily extended to the more general case of arbitrary number-states $|n_1\rangle$ and $|n_2\rangle$ arriving simultaneously at the entrance ports of an RBS.[9])

The basic properties of a certain type of polarizing beam-splitter (the birefringent type) and also those of wave-plates (birefringent or otherwise) are briefly reviewed in Sec.3.[10,11] Then, in Secs. 4 and 5, we examine several cases of single-mode, 2-photon states that propagate along the *z*-axis of a Cartesian coordinate system, are linearly-polarized at 45° to the $x$ and $y$ axes, and pass through a polarizing beam-splitter (PBS), or a quarter-wave plate, or a half-wave plate. A similar analysis of single-mode, circularly-polarized, 2-photon states that pass through a PBS, or a half-wave plate, or a quarter-wave plate, is the subject of Sec.6. Several examples at the end of Sec.6 extend and further elaborate the results of earlier sections. It will be seen through these examples that, while the operator algebraic techniques may be quick and pain-free, the combinatoric methods of counting the alternative paths taken by the photons to their destinations are equally valid and far more insightful.

A different application of the Feynman principle, involving a continuum of indistinguishable paths to a given observable event, is presented in Sec.7. Here, we show how a single photon impinging on a paraboloidal mirror converges to a small region within an observation plane, with an amplitude profile that is analogous to the conventional Airy pattern observed at the focus of lenses and mirrors in classical physical optics.[10-14] The paper ends with a few concluding remarks in Sec.8.

**2. Photon number redistribution (and Bose enhancement) at a regular beam-splitter**. Consider the lossless, symmetric, regular beam-splitter (RBS) depicted in Fig.1, having reflection coefficient $\rho = |\rho|e^{i\varphi}$ and transmission coefficient $\tau = |\tau|e^{i(\varphi+\frac{1}{2}\pi)}$ at both its entrance ports 1 and 2, with $|\rho|^2 + |\tau|^2 = 1$. (See the Appendix for a discussion of the 90° phase difference between $\rho$ and $\tau$.) A wavepacket in the (nearly) single mode $(\boldsymbol{k}, \omega, \hat{\boldsymbol{e}})$, with $\boldsymbol{k} = (\omega/c)\hat{\boldsymbol{z}}$ and linear polarization $\hat{\boldsymbol{e}} = \hat{\boldsymbol{y}}$, carrying two photons in the number-state $|2\rangle$, arrives at the splitter's port 1. At the same time, a second (also nearly single-mode) packet, this one with $\boldsymbol{k} = (\omega/c)\hat{\boldsymbol{x}}$ and $\hat{\boldsymbol{e}} = \hat{\boldsymbol{y}}$, carrying only one photon in the number-state $|1\rangle$, arrives at the splitter's port 2. One may imagine the individual photons within these packets as being point-particles (or tiny little balls) that travel with the speed of light $c$ along the direction of their corresponding $k$-vector. Each photon is located somewhere within its host wavepacket, although its precise spacetime coordinates remain unknown until the photon is detected — say, by a large-area, high-resolution, single-photon-detecting array of avalanche photo-multipliers. If our three photons were not identical, one could have imagined them as three little colored balls, each with its own distinctive color, traveling toward one or the other entrance port of the RBS. However, since the prevailing assumption here is that the photons are indistinguishable, we must take the three hypothetical little balls to be identical in every respect.

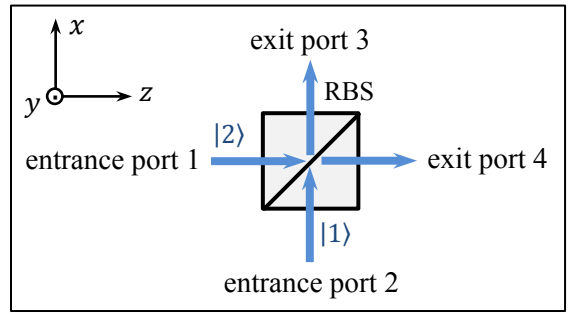

**Fig.1**. Arriving at a regular beam-splitter are three photons: two through the entrance port 1, and one through port 2. The three photons are subsequently redistributed between the exit ports 3 and 4.

Let a pair of single-photon detectors reside at the exit ports 3 and 4 of the RBS; these detectors monitor the arrival times of individual photons at the two exit ports. Taking the overall duration of each incoming packet to be $T$, we divide this interval into a large number $N$ of short time windows $\Delta t = T/N$, within which single photodetection events can occur.[†] We assume that each and every photon-capture event by either detector results in a click that can be distinctly monitored and recorded. Each click is then associated with a single-photon detection time $t_n = (n + ½)\Delta t$ at the mid-point of one of the $N$ available time windows at either one of the exit ports. (The slight chance of two or more photons arriving at a given detector within the same time window can be ignored, provided that the total number $N$ of available time slots is sufficiently large.)

---

[†] In principle, each exit port should be equipped with a large, pixelated array of single-photon detectors, so that the captured photons are localized (as accurately as possible) in space as well as in time. This, in fact, is what is demanded by the strict notion of observable events and the distinguishability or indistinguishability of such events, which, in the system presently under consideration, is associated with the "clicks" heard at individual detectors. It can be shown,[9] however, that the end result of the analysis will be the same, irrespective of whether the single photodetection events are localized in time alone, or in space as well as in time. What matters most is that no two clicks should coincide.

The probability amplitude that all three incoming photons are directed toward port 3 is $\rho^2 \tau$, in which case three clicks will be heard at the detector located in port 3, say, at times $t_i, t_j, t_k$. These three clicks, within the specified time windows, occur with a probability of $1/N^3$, or a probability amplitude of $1/\sqrt{N^3}$. Had the three photons been distinct, there would have been 3! permutations of them arriving at $t_i, t_j, t_k$, but their indistinguishability dictates that we add up these amplitudes to arrive at $3!/\sqrt{N^3}$. Since there are $N$ available time windows in port 3 and the clicks can occur in any three of them, the total (possible) number of such 3-photon detection events is going to be

$$\binom{N}{3} = \frac{N!}{3!(N-3)!} \cong \frac{N^3}{3!}. \tag{1}$$

These events, however, are all distinguishable from one another (since they occur in different time slots), and one must add up their probabilities — rather than their amplitudes. Thus, the overall amplitude that one would hear three clicks at port 3 amounts to $(3!/\sqrt{N^3})\sqrt{N^3/3!}\,\rho^2\tau = \sqrt{3!}\,\rho^2\tau$. (Note that we have multiplied here the individual event amplitude by $\sqrt{N^3/3!}$, which is the same as multiplying the corresponding probability by $N^3/3!$.)

Suppose now that both photons that enter through port 1 go to port 3, while the single photon in port 2 is directed toward port 4; the amplitude for this occurrence is $\rho^3$. The two clicks in port 3 have a probability of $1/N^2$ (or amplitude of $1/N$) to be heard at specific instances, say, $t_i$ and $t_j$. These clicks can occur by two different permutations of the photons heading from port 1 to port 3, and since these two photons are indistinguishable, the corresponding amplitudes must be added together, yielding $2!/N$. Considering that the total number of distinct two-click events in port 3 is

$$\binom{N}{2} = \frac{N!}{2!(N-2)!} \cong \frac{N^2}{2!}, \tag{2}$$

the resulting amplitude for the event under consideration turns out to be $(2!/N)\sqrt{N^2/2!}\,\rho^3 = \sqrt{2!}\,\rho^3$.

An alternative way of getting two photons in port 3 (and one in port 4) is to let one photon from port 1 go to port 3 and the other one to port 4; the photon entering through port 2 must then end up in port 3 as well. The amplitude for this occurrence is $\rho\tau^2$. Since there are two photons in port 1 and either one could be the one that ends up in port 3, we have two options here. We also have two ways to permutate the photons in port 3 that could bring about clicks at times $t_i$ and $t_j$. Considering that the number of available pairs of time slots in port 3 is the same as that given by Eq.(2), the overall amplitude of the particular event under consideration will be $2(2!/N)\sqrt{N^2/2!}\,\rho\tau^2 = 2\sqrt{2!}\,\rho\tau^2$. It is now seen that two different, albeit indistinguishable, paths can lead to the same output state $|2\rangle_3|1\rangle_4$; adding up the corresponding amplitudes, we arrive at $\sqrt{2!}\,(\rho^3 + 2\rho\tau^2)$.

Similar arguments can be advanced for the remaining output states $|1\rangle_3|2\rangle_4$ and $|0\rangle_3|3\rangle_4$. In fact, the amplitudes of these states can be obtained from those already computed by exchanging $\rho$ and $\tau$. All in all, the emergent superposition state at the exit ports 3 and 4 of the RBS is found to be

$$\sqrt{3!}\,\rho^2\tau|3\rangle_3|0\rangle_4 + \sqrt{2!}\,(\rho^3 + 2\rho\tau^2)|2\rangle_3|1\rangle_4 + \sqrt{2!}\,(\tau^3 + 2\tau\rho^2)|1\rangle_3|2\rangle_4 + \sqrt{3!}\,\tau^2\rho|0\rangle_3|3\rangle_4. \tag{3}$$

The above state still needs to be normalized, since the sum of its probabilities (i.e., the squared magnitudes of the amplitudes) is greater than 1, as seen below:

$$\text{sum of probabilities} = 6|\rho|^4|\tau|^2 + 2|\rho|^6 + 8|\rho|^2|\tau|^4 + 4|\rho|^2\rho^2\tau^{*2} + 4|\rho|^2\rho^{*2}\tau^2$$
$$+ 2|\tau|^6 + 8|\rho|^4|\tau|^2 + 4|\tau|^2\tau^2\rho^{*2} + 4|\tau|^2\tau^{*2}\rho^2 + 6|\rho|^2|\tau|^4$$
$$= 2|\rho|^6 + 6|\rho|^4|\tau|^2 + 6|\rho|^2|\tau|^4 + 2|\tau|^6 = 2(|\rho|^2 + |\tau|^2)^3 = 2. \tag{4}$$

Dividing the state of Eq.(3) by $\sqrt{2}$, we finally arrive at the normalized emergent state, namely,

$$\sqrt{3}\rho^2\tau|3\rangle_3|0\rangle_4 + (\rho^3 + 2\rho\tau^2)|2\rangle_3|1\rangle_4 + (\tau^3 + 2\tau\rho^2)|1\rangle_3|2\rangle_4 + \sqrt{3}\tau^2\rho|0\rangle_3|3\rangle_4. \tag{5}$$

The need for the above normalization by $\sqrt{2}$ can be traced back to the annihilation of the input state $|2\rangle_1|1\rangle_2$, since $\hat{a}_1^2\hat{a}_2|2\rangle_1|1\rangle_2 = \sqrt{2!}\,|0\rangle|0\rangle$. The annihilation of an incoming number-state $|n\rangle$ at the beam-splitter will always require a division of the final amplitudes by $\sqrt{n!}$. In contrast, the creation of a number-state $|n\rangle$ from vacuum introduces an enhancement factor of $\sqrt{n!}$. This explains why the first and fourth terms in Eq.(5) have a Bose enhancement factor of $\sqrt{3}$: first a 2-photon state is annihilated in port 1, then a 3-photon state is created in either port 3 or port 4. In contrast, when the annihilation of a 2-photon state is followed by the creation of another 2-photon state, there will be no Bose enhancement factors, which is why $\rho^3$ in the second term and $\tau^3$ in the third term of Eq.(5) are not enhanced. It goes without saying that the enhancement factors of 2 that accompany $\rho\tau^2$ and $\rho^2\tau$ in the second and third terms of Eq.(5) arise from the fact that two possibilities exist for one of the two photons from port 1 to go to port 3 while the other one ends up in port 4.

In the remainder of this section, we show that the emergent state of Eq.(5) can be obtained directly via the standard operator algebra of quantum electrodynamics.[6-8] The annihilation operators at the exit ports 3 and 4 of a lossless and symmetric beam-splitter are related to those at its entrance ports 1 and 2, as follows:

$$\hat{a}_3 = \rho\hat{a}_1 + \tau\hat{a}_2 \qquad \text{and} \qquad \hat{a}_4 = \tau\hat{a}_1 + \rho\hat{a}_2. \tag{6}$$

Considering that the annihilation and creation operators for the states arriving at ports 1 and 2 are fully independent of each other, it should not be surprising to learn that $\hat{a}_1$ commutes with $\hat{a}_2$; that is, $[\hat{a}_1, \hat{a}_2] = 0$. It is now easy to verify, with the aid of Eqs.(6), that $\hat{a}_3$ and $\hat{a}_4$ also commute with each other; that is, $[\hat{a}_3, \hat{a}_4] = 0$.

Expressing $\hat{a}_1$ and $\hat{a}_2$ in terms of $\hat{a}_3$ and $\hat{a}_4$ requires the inversion of Eqs.(6), which starts with the following evaluation of the determinant of the coefficients matrix:

$$\rho^2 - \tau^2 = |\rho|^2 e^{2i\varphi} - |\tau|^2 e^{2i(\varphi+\frac{1}{2}\pi)} = (|\rho|^2 + |\tau|^2)e^{2i\varphi} = e^{2i\varphi}. \tag{7}$$

Inversion of Eqs.(6) now yields

$$\begin{pmatrix}\hat{a}_1\\\hat{a}_2\end{pmatrix} = \begin{pmatrix}\rho & \tau\\\tau & \rho\end{pmatrix}^{-1}\begin{pmatrix}\hat{a}_3\\\hat{a}_4\end{pmatrix} = (\rho^2-\tau^2)^{-1}\begin{pmatrix}\rho & -\tau\\-\tau & \rho\end{pmatrix}\begin{pmatrix}\hat{a}_3\\\hat{a}_4\end{pmatrix}$$
$$= e^{-2i\varphi}\begin{pmatrix}|\rho|e^{i\varphi} & -|\tau|e^{i(\varphi+\frac{1}{2}\pi)}\\-|\tau|e^{i(\varphi+\frac{1}{2}\pi)} & |\rho|e^{i\varphi}\end{pmatrix}\begin{pmatrix}\hat{a}_3\\\hat{a}_4\end{pmatrix} = \begin{pmatrix}\rho^* & \tau^*\\\tau^* & \rho^*\end{pmatrix}\begin{pmatrix}\hat{a}_3\\\hat{a}_4\end{pmatrix}. \tag{8}$$

Consequently,

$$\hat{a}_1 = \rho^*\hat{a}_3 + \tau^*\hat{a}_4 \qquad \text{and} \qquad \hat{a}_2 = \tau^*\hat{a}_3 + \rho^*\hat{a}_4. \tag{9}$$

The transpose conjugates of the above equations are

$$\hat{a}_1^\dagger = \rho\hat{a}_3^\dagger + \tau\hat{a}_4^\dagger \qquad \text{and} \qquad \hat{a}_2^\dagger = \tau\hat{a}_3^\dagger + \rho\hat{a}_4^\dagger. \tag{10}$$

We are finally in a position to relate the input state at the entrance ports of the RBS to the emergent state at its output ports, as follows:

$$|2\rangle_1|1\rangle_2 = (\hat{a}_1^\dagger)^2\hat{a}_2^\dagger|0\rangle|0\rangle = (\rho\hat{a}_3^\dagger + \tau\hat{a}_4^\dagger)^2(\tau\hat{a}_3^\dagger + \rho\hat{a}_4^\dagger)|0\rangle|0\rangle$$

$$= [\rho^2\tau(\hat{a}_3^\dagger)^3 + (\rho^3 + 2\rho\tau^2)(\hat{a}_3^\dagger)^2\hat{a}_4^\dagger + (\tau^3 + 2\rho^2\tau)\hat{a}_3^\dagger(\hat{a}_4^\dagger)^2 + \rho\tau^2(\hat{a}_4^\dagger)^3]|0\rangle|0\rangle$$

$$= \sqrt{3!}\,\rho^2\tau|3\rangle_3|0\rangle_4 + \sqrt{2!}\,(\rho^3 + 2\rho\tau^2)|2\rangle_3|1\rangle_4 + \sqrt{2!}\,(\tau^3 + 2\rho^2\tau)|1\rangle_3|2\rangle_4 + \sqrt{3!}\,\rho\tau^2|0\rangle_3|3\rangle_4. \quad (11)$$

The state in Eq.(11) is the same as that in Eq.(3), which, upon normalization (i.e., division by $\sqrt{2}$) coincides with the state in Eq.(4). The need for normalization, of course, arises from the fact that, in the first line of Eq.(11), we deliberately left off a factor of $\sqrt{2!}$ when equating $(\hat{a}_1^\dagger)^2\hat{a}_2^\dagger|0\rangle|0\rangle$ with $|2\rangle_1|1\rangle_2$ instead of $\sqrt{2!}\,|2\rangle_1|1\rangle_2$. This was meant to emphasize that the need for normalization traces back to the same root cause, irrespective of whether one relies on the operator algebra or uses the combinatoric approach to compute the probability amplitudes at the exit ports.[9]

**3. Polarizing beam-splitters and wave-plates**. The essential features of a polarizing beam-splitter (PBS) are shown in Fig.2(a). When a single-mode wavepacket $(\boldsymbol{k}, \omega, \hat{\boldsymbol{e}})$ enters through port 1 of the splitter, the emergent packets leave the PBS through ports 3 and 4. Suppose the first prism, shown on the left-hand side of Fig.2(a), is made of an isotropic material of refractive index $n_0$, while the second prism's material is biaxially birefringent,[10,11] having refractive indices $(n_x, n_y, n_z)$ along the $x, y, z$ coordinate axes. Since the $x$-polarized light arriving in port 1 must be perfectly transmitted to port 4, we set $n_x = n_z = n_0$ to ensure that no $x$-polarized light gets reflected at the interface between the two prisms. As for the $y$-polarized incident light, we ask that total internal reflection take place at the interface,[10-14] which requires not only that $n_y$ be less than $n_0$, but also that the incidence angle (45° in the diagram) be greater than the critical angle $\theta_c = \sin^{-1}(n_y/n_0)$ of total internal reflection.[‡] Under the circumstances, the phase difference between the reflected and transmitted waves would depend on the values of the refractive indices and the angle of incidence at the interface. Thus, in contrast to the case of a non-polarizing, lossless beam-splitter, where energy conservation imposes a 90° phase difference between the reflection and transmission coefficients, in the case of a PBS, the relative phase between the reflected and transmitted beams is not constrained. We proceed, without loss of generality, to assume that the emergent beams at ports 3 and 4 of the PBS are in-phase.

Figure 2(b) shows a wave-plate that transmits with zero phase shift an incident electromagnetic wave polarized along the plate's fast axis $\hat{\boldsymbol{x}}$, while imparting a phase $\varphi$ to an incident beam that is polarized along the plate's slow axis $\hat{\boldsymbol{y}}$. For quarter-wave plates, $\varphi = 90°$, whereas for a half-wave plate, $\varphi = 180°$.[10,11] If a beam in the number-state $|n\rangle_y$, containing $n$ identical photons polarized along $\hat{\boldsymbol{y}}$, passes through the plate, its probability amplitude will get multiplied by $e^{in\varphi}$. Indeed, this is needed to ensure that an incident $y$-polarized coherent beam $|\gamma\rangle_y = e^{-|\gamma|^2/2}\sum_{n=0}^{\infty}(\gamma^n/\sqrt{n!})|n\rangle_y$ emerges from the plate as the properly phase-shifted coherent beam $|\gamma e^{i\varphi}\rangle_y$.[4,6-8]

---

[‡] The birefringent PBS of Fig.2(a) fails to act as a good PBS if the incident beam arrives through port 2. It is true that an incoming $z$-polarized beam at port 2 will be fully transmitted to port 3, simply because $n_x = n_z = n_0$. However, recalling that $n_y$ is less than $n_0$, it is readily seen that the $y$-polarized incident beam in port 2 is only partially reflected (toward port 4) at the interface between the two prisms. Moreover, the $k$-vector of the transmitted portion of the $y$-polarized light that emerges from port 3 will deviate from the $x$-axis. If one desires to operate the PBS of Fig.2(a) with two incident beams that arrive simultaneously in ports 1 and 2, one must coat the interface between the prisms with a birefringent multilayer stack that would provide full reflectivity for the $y$-polarized light entering through port 2, without adversely affecting the remaining reflection and transmission coefficients of the PBS.

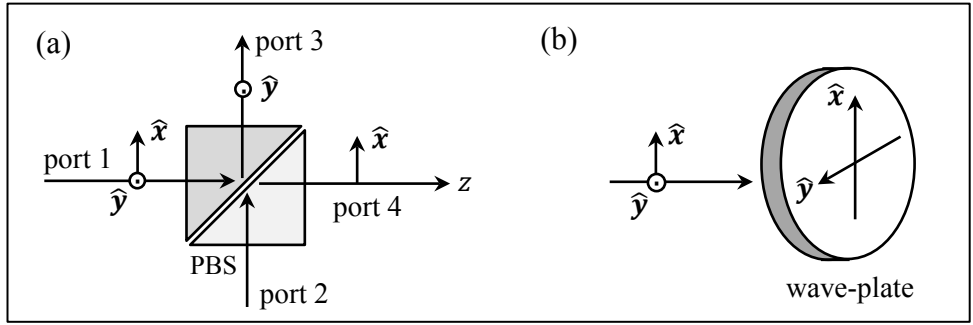

**Fig.2**. (a) Polarizing beam-splitter (PBS) with its transmission and reflection axes aligned with the unit-vectors $\hat{x}$ and $\hat{y}$ of a Cartesian coordinate system. When a single-mode wavepacket $(k, \omega, \hat{e})$ enters through port 1, the emergent packets leave the PBS through ports 3 and 4. Since there is no mandatory phase relation between the reflected and transmitted waves, the relative phase of the emergent beams may be taken to be zero. The birefringent type of PBS described in Sec.3 does *not* behave properly for an incident beam arriving through port 2, unless a multilayer birefringent coating that fully reflects the $y$-polarization (while continuing to transmit the $z$-polarization) is interposed between the two prisms. (b) In classical optics, a wave-plate imparts a phase $\varphi$ to the incident $E$-field along its slow axis $\hat{y}$, while the incident $E$-field along the fast axis $\hat{x}$ emerges with zero phase shift.

**4. Creation and annihilation operators for polarized photons in various number-states**. The $E$-field operator for a propagating single-mode $(k, \omega, \hat{e})$ electromagnetic wave in free space, whose polarization is specified by the complex unit-vector $\hat{e}$ (that is, $\hat{e} \cdot \hat{e}^* = 1$), is known to be[6-8]

$$\widehat{E}(r,t) = i\sqrt{\hbar\omega/(2\varepsilon_0 V)}\left[e^{i(k\cdot r - \omega t)}\hat{e}\hat{a} - e^{-i(k\cdot r - \omega t)}\hat{e}^*\hat{a}^\dagger\right]. \tag{12}$$

Here, $\hbar\omega$ is the energy of a single photon, $\varepsilon_0$ is the permittivity of free space, and $V$ is the (large) spatial volume occupied by the electromagnetic field under consideration. Equation (12) intimates that $\hat{e}^*\hat{a}^\dagger$ is the creation operator for a single photon in the polarization state $\hat{e}$. For linear polarization at 45° to the $x$-axis, $\hat{e} = (\hat{x} + \hat{y})/\sqrt{2}$, whereby a single-photon creation operation yields

$$\tfrac{1}{\sqrt{2}}(\hat{a}_x^\dagger + \hat{a}_y^\dagger)|0\rangle_x|0\rangle_y = \tfrac{1}{\sqrt{2}}|1\rangle_x|0\rangle_y + \tfrac{1}{\sqrt{2}}|0\rangle_x|1\rangle_y. \tag{13}$$

Similarly, invoking the two-photon creation operator for linear polarization at 45° to $\hat{x}$, we find

$$\tfrac{1}{2}(\hat{a}_x^\dagger + \hat{a}_y^\dagger)(\hat{a}_x^\dagger + \hat{a}_y^\dagger)|0\rangle_x|0\rangle_y = \tfrac{1}{2}\left(\hat{a}_x^{\dagger 2} + \hat{a}_y^{\dagger 2} + 2\hat{a}_x^\dagger\hat{a}_y^\dagger\right)|0\rangle_x|0\rangle_y$$

$$= \tfrac{\sqrt{2}}{2}|2\rangle_x|0\rangle_y + \tfrac{\sqrt{2}}{2}|0\rangle_x|2\rangle_y + |1\rangle_x|1\rangle_y. \quad \boxed{\text{divide by } \sqrt{2} \text{ to normalize}} \tag{14}$$

Upon normalization (i.e., division by $\sqrt{2}$), the state given by Eq.(14) will be a 2-photon state polarized at 45° to $\hat{x}$. If a wavepacket in this state enters a PBS whose transmission and reflection axes are aligned with $\hat{x}$ and $\hat{y}$, it would be tempting to suppose that the emergent states will be

$$\text{port 3:} \quad \tfrac{1}{2}|0\rangle_y + \tfrac{1}{\sqrt{2}}|1\rangle_y + \tfrac{1}{2}|2\rangle_y, \tag{15}$$

$$\text{port 4:} \quad \tfrac{1}{2}|0\rangle_x + \tfrac{1}{\sqrt{2}}|1\rangle_x + \tfrac{1}{2}|2\rangle_x. \tag{16}$$

This, however, is *not* correct, since the emergent packets in ports 3 and 4 are entangled. It is true that the probability of catching zero, one, or two $y$-polarized photons in port 3 will be ¼, ½, and ¼, respectively, and that similar probabilities govern the emergence of zero, one, or two $x$-polarized photons at port 4. Nevertheless, once the measurement at one exit port has ascertained the number of photons emerging from that port, there shall be no uncertainty as to the number (and polarization state) of those that arrive at the other exit port.

Let us now consider sending a wavepacket in the state of Eq.(14) (after normalization) through a PBS whose transmission and reflection axes $\hat{x}'$ and $\hat{y}'$ are rotated around the $z$-axis by $45°$ away from $\hat{x}$ and $\hat{y}$. The probability amplitude that both photons will come out of port 4 (i.e., polarized along $\hat{x}'$) is readily found to be

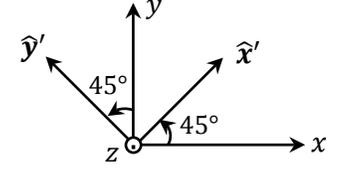

$$\tfrac{1}{2}\left(\tfrac{1}{\sqrt{2}}\right)^2 + \tfrac{1}{2}\left(\tfrac{1}{\sqrt{2}}\right)^2 + \tfrac{1}{\sqrt{2}}\left(\tfrac{1}{\sqrt{2}}\right)\left(\tfrac{1}{\sqrt{2}}\right)\sqrt{2} = ¼ + ¼ + ½ = 1. \quad (17)$$

[Bose enhancement factor]

The sum of the three amplitudes on the left-hand side of the above equation corresponds to three indistinguishable possibilities, namely, that (i) both $x$-polarized photons emerge with their polarization along $\hat{x}'$; (ii) both $y$-polarized photons emerge with their polarization along $\hat{x}'$; (iii) the single $x$-polarized photon and also the single $y$-polarized photon emerge with their polarization along $\hat{x}'$. Similarly, the amplitude of both photons emerging from port 3 (i.e., polarized long $\hat{y}'$) is

[Bose enhancement factor]

$$\tfrac{1}{2}\left(-\tfrac{1}{\sqrt{2}}\right)^2 + \tfrac{1}{2}\left(\tfrac{1}{\sqrt{2}}\right)^2 + \tfrac{1}{\sqrt{2}}\left(-\tfrac{1}{\sqrt{2}}\right)\left(\tfrac{1}{\sqrt{2}}\right)\sqrt{2} = ¼ + ¼ - ½ = 0. \quad (18)$$

Finally, the amplitude that one photon will come out of port 3 and the other out of port 4 is

[$\sqrt{2}$ factors are needed here, because each 2-photon state can split between $\hat{x}'$ and $\hat{y}'$ in two different ways]

$$\tfrac{1}{2}\left(\tfrac{1}{\sqrt{2}}\right)\left(-\tfrac{1}{\sqrt{2}}\right)\sqrt{2} + \tfrac{1}{2}\left(\tfrac{1}{\sqrt{2}}\right)\left(\tfrac{1}{\sqrt{2}}\right)\sqrt{2} + \tfrac{1}{\sqrt{2}}\left[\left(\tfrac{1}{\sqrt{2}}\right)\left(\tfrac{1}{\sqrt{2}}\right) + \left(-\tfrac{1}{\sqrt{2}}\right)\left(\tfrac{1}{\sqrt{2}}\right)\right] = 0. \quad (19)$$

As expected, both photons in this case emerge from port 4 of the rotated PBS.

**5. Two-photon wavepacket linearly-polarized at $45°$ to $x$-axis passing through a wave-plate.** If the (normalized) 2-photon packet of Eq.(14) enters a quarter-wave plate (also known as a $\lambda/4$-plate) whose axes are aligned with $\hat{x}$ and $\hat{y}$, the emerging state will be

[two-photon phase shift $e^{i\pi} = -1$] [one-photon phase shift $e^{i\pi/2} = i$]

$$|\psi\rangle = \tfrac{1}{2}|2\rangle_x|0\rangle_y - \tfrac{1}{2}|0\rangle_x|2\rangle_y + \tfrac{i}{\sqrt{2}}|1\rangle_x|1\rangle_y. \quad (20)$$

As expected, this is the same as a right-circularly-polarized (RCP) two-photon state, namely,

$$|\psi\rangle = ½(\hat{a}_x^\dagger + i\hat{a}_y^\dagger)(\hat{a}_x^\dagger + i\hat{a}_y^\dagger)|0\rangle_x|0\rangle_y = ½\big(\hat{a}_x^{\dagger 2} - \hat{a}_y^{\dagger 2} + i2\hat{a}_x^\dagger\hat{a}_y^\dagger\big)|0\rangle_x|0\rangle_y$$

$$= \tfrac{\sqrt{2}}{2}|2\rangle_x|0\rangle_y - \tfrac{\sqrt{2}}{2}|0\rangle_x|2\rangle_y + i|1\rangle_x|1\rangle_y. \quad \leftarrow \text{[divide by } \sqrt{2} \text{ to normalize]} \quad (21)$$

In contrast, if the 2-photon packet that is linearly-polarized at $45°$ to the $x$-axis goes through a half-wave (i.e., $\lambda/2$) plate whose axes are aligned with $\hat{x}$ and $\hat{y}$, the emergent state will be

[two-photon phase shift $e^{i2\pi} = 1$] [one-photon phase shift $e^{i\pi} = -1$]

$$|\psi\rangle = \tfrac{1}{2}|2\rangle_x|0\rangle_y + \tfrac{1}{2}|0\rangle_x|2\rangle_y - \tfrac{1}{\sqrt{2}}|1\rangle_x|1\rangle_y. \quad (22)$$

The emergent polarization is expected to be linear once again but rotated by $90°$; that is, it must be at $-45°$ relative to $\hat{x}$. Recall that, for linear polarization at $-45°$ to the $x$-axis, $\hat{e} = (\hat{x} - \hat{y})/\sqrt{2}$. Consequently, the single-photon and two-photon states created in this state of polarization are

$$|\psi\rangle = \tfrac{1}{\sqrt{2}}(\hat{a}_x^\dagger - \hat{a}_y^\dagger)|0\rangle_x|0\rangle_y = \tfrac{1}{\sqrt{2}}|1\rangle_x|0\rangle_y - \tfrac{1}{\sqrt{2}}|0\rangle_x|1\rangle_y. \quad (23)$$

$$|\psi\rangle = \tfrac{1}{2}(\hat{a}_x^\dagger - \hat{a}_y^\dagger)(\hat{a}_x^\dagger - \hat{a}_y^\dagger)|0\rangle_x|0\rangle_y = \tfrac{1}{2}(\hat{a}_x^{\dagger 2} + \hat{a}_y^{\dagger 2} - 2\hat{a}_x^\dagger \hat{a}_y^\dagger)|0\rangle_x|0\rangle_y$$

$$= \tfrac{\sqrt{2}}{2}|2\rangle_x|0\rangle_y + \tfrac{\sqrt{2}}{2}|0\rangle_x|2\rangle_y - |1\rangle_x|1\rangle_y. \quad \leftarrow \boxed{\text{divide by } \sqrt{2} \text{ to normalize}} \quad (24)$$

The linearly-polarized 2-photon state of Eq.(24) is clearly the same as that given by Eq.(22).

**6. Circularly polarized one-photon and two-photon states**. Single-photon states with right- and left-circular polarization, namely, $\hat{e} = (\hat{x} \pm i\hat{y})/\sqrt{2}$, are created by the following operations:

$$\tfrac{1}{\sqrt{2}}(\hat{a}_x^\dagger \pm i\hat{a}_y^\dagger)|0\rangle_x|0\rangle_y = \tfrac{1}{\sqrt{2}}|1\rangle_x|0\rangle_y \pm \tfrac{i}{\sqrt{2}}|0\rangle_x|1\rangle_y. \quad (25)$$

A two-photon state, created with one RCP and one LCP photon, is

$$|\psi\rangle = \tfrac{1}{2}(\hat{a}_x^\dagger + i\hat{a}_y^\dagger)(\hat{a}_x^\dagger - i\hat{a}_y^\dagger)|0\rangle_x|0\rangle_y = \tfrac{1}{2}(\hat{a}_x^{\dagger 2} + \hat{a}_y^{\dagger 2})|0\rangle_x|0\rangle_y = \tfrac{\sqrt{2}}{2}|2\rangle_x|0\rangle_y + \tfrac{\sqrt{2}}{2}|0\rangle_x|2\rangle_y. \quad (26)$$

In contrast to the 2-photon state of Eq.(14), or Eq.(24), which needed normalization (by a factor of $\sqrt{2}$), the 2-photon state of Eq.(26) is already normalized in its current form; the underlying reason is that the two photons in Eq.(14), or Eq.(24), were identical, whereas those in Eq.(26) are not.

If the wavepacket of Eq.(26) goes through a PBS whose axes are aligned with $\hat{x}$ and $\hat{y}$, both photons will emerge from port 3 with probability ½, or both from port 4, also with probability ½.

If the wavepacket of Eq.(26) goes through a half-wave plate whose axes are aligned with $\hat{x}$ and $\hat{y}$, the amplitude of its $y$-polarized component gets multiplied by $e^{i2\pi} = 1$; therefore, the emergent state remains unchanged, consistent with the fact that the $\lambda/2$-plate converts the incident RCP photon to LCP, and vice-versa. Clearly, the incident packet in the present example, containing one RCP and one LCP photon, is unaffected by its passage through a $\lambda/2$-plate.

If the wavepacket of Eq.(26) passes through a $\lambda/4$-plate whose axes are aligned with $\hat{x}$ and $\hat{y}$, the amplitude of its $y$-polarized component gets multiplied by $e^{i\pi} = -1$, leading to the following emergent state:

$$|\psi\rangle = \tfrac{1}{\sqrt{2}}|2\rangle_x|0\rangle_y - \tfrac{1}{\sqrt{2}}|0\rangle_x|2\rangle_y. \quad (27)$$

This is an equal superposition of a pair of 2-photon states, one in which both photons are RCP, such as that of Eq.(21), the other, a similar state with both photons being LCP. Note that the incident packet of Eq.(26) can be considered a superposition of two linearly-polarized, 2-photon states, one polarized at 45° to the $x$-axis, such as that of Eq.(14), the other at −45°, such as that of Eq.(24). Consequently, the state in Eq.(27) that emerges from the $\lambda/4$-plate is the result of the conversion of two mutually orthogonal linear polarization states to a pair of RCP and LCP states.

The above interpretation of Eq.(27) may, at first glance, seem a bit counterintuitive, as one might have expected the $\lambda/4$-plate to turn the incoming packet of one RCP and one LCP photon into a packet containing two linearly-polarized photons whose respective polarizations are oriented at ±45° to the $x$-axis. However, there is no discrepancy here, since an alternative interpretation of Eq.(27) is that $|\psi\rangle$ is the result of cascade creation of a pair of linearly-polarized photons at ±45° to the $x$-axis, namely, $\tfrac{1}{2}(\hat{a}_x^\dagger - \hat{a}_y^\dagger)(\hat{a}_x^\dagger + \hat{a}_y^\dagger)|0\rangle_x|0\rangle_y$, which is precisely the state appearing in Eq.(27).

**Example 1**. Let the 2-photon state $|2\rangle_y$ arrive at a $\lambda/2$-plate whose fast and slow axes, $\hat{x}'$ and $\hat{y}'$, are rotated by 45° around the $z$-axis. The amplitude that both photons experience the plate's fast axis is $(1/\sqrt{2})^2 = ½$, in which case they emerge with polarization along the $\hat{x}'$ axis of the plate. There is the same ½ amplitude that both photons will experience the plate's slow axis and emerge with a phase $e^{i2\pi} = 1$ and polarization aligned with the $\hat{y}'$ axis. The third and final possibility is for one photon to experience the fast axis and the other the slow axis, in which case the slow axis will impart a phase shift $e^{i\pi} = -1$ to the emergent state. The amplitude of this event, however, is $(1/\sqrt{2})^2 e^{i\pi}\sqrt{2} = -1/\sqrt{2}$, which includes a $\sqrt{2}$ enhancement factor. This is because the initial breakup of the 2-photon state calls for a division by 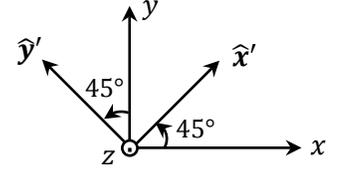
$\sqrt{2!}$, which is then followed by a multiplication by 2, as there are two indistinguishable ways in which one photon could experience the plate's fast axis while the other one interacts with the slow axis. All in all, the emergent state is given by

$$|\psi\rangle = \tfrac{1}{2}|2\rangle_{x'}|0\rangle_{y'} + \tfrac{1}{2}|0\rangle_{x'}|2\rangle_{y'} - \tfrac{1}{\sqrt{2}}|1\rangle_{x'}|1\rangle_{y'}. \tag{28}$$

The emergent 2-photon state is seen to be linearly-polarized at $-45°$ to the $x'$-axis; see Eq.(24). Thus, the emergent wavepacket is linearly-polarized along the original $x$-axis, which confirms that the $\lambda/2$-plate has rotated the polarization of the incident photon pair by 90° around the $z$-axis. In this way, we have constructed the 2-photon state of Eq.(28) with the aid of the Feynman method, without any explicit or implicit resorts to the operator algebra, which was used to arrive at Eq.(24).

**Example 2**. Let the 2-photon state $|2\rangle_y$ arrive at a $\lambda/4$-plate whose fast and slow axes, $\hat{x}'$ and $\hat{y}'$, are rotated by 45° around the $z$-axis. The amplitude that both photons experience the plate's fast axis is $(1/\sqrt{2})^2 = ½$, in which case they emerge with polarization along the $\hat{x}'$ axis. There is the same ½ amplitude that both photons will experience the plate's slow axis and emerge with a phase $e^{i\pi} = -1$ and polarization aligned with the $\hat{y}'$ axis. The third and final possibility is for one photon to experience the fast axis and the other the slow axis, in 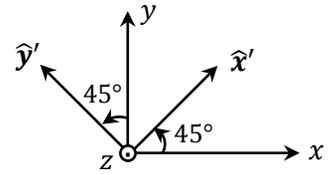
which case the slow axis will impart a phase shift $e^{i\pi/2} = i$ to the emergent state. The amplitude of this event, however, is $(1/\sqrt{2})^2 e^{i\pi/2}\sqrt{2} = i/\sqrt{2}$, which includes the $\sqrt{2}$ enhancement factor. As in the preceding example, the enhancement factor comes about because the initial breakup of the 2-photon state calls for a division by $\sqrt{2!}$, which is then followed by a multiplication by 2, since there are two indistinguishable ways in which one photon could experience the plate's fast axis and the other one the slow axis. The emergent state is, therefore,

$$|\psi\rangle = \tfrac{1}{2}|2\rangle_{x'}|0\rangle_{y'} - \tfrac{1}{2}|0\rangle_{x'}|2\rangle_{y'} + \tfrac{i}{\sqrt{2}}|1\rangle_{x'}|1\rangle_{y'}. \tag{29}$$

The emergent 2-photon state is what one obtains by applying the creation operator for two RCP photons to the vacuum state $|0\rangle_{x'}|0\rangle_{y'}$; see Eq.(21). This confirms that the $\lambda/4$-plate has converted the pair of incident photons from linear to circular polarization.

**Example 3**. A single-mode 2-photon wavepacket is linearly-polarized at 45° to the $x$-axis. We examine the emerging state from port 4 of a PBS whose transmission axis $\hat{x}'$ is at an angle $\theta$ relative to the $x$-axis. With the incident state being $|\psi\rangle = ½|2\rangle_x|0\rangle_y + ½|0\rangle_x|2\rangle_y + (1/\sqrt{2})|1\rangle_x|1\rangle_y$, the probability amplitudes that zero, one, or two photons emerge from port 4 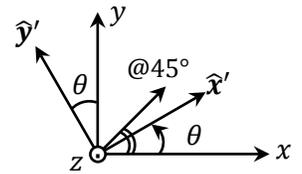

of the PBS (with transmission axis $\hat{x}'$ and reflection axis $\hat{y}'$) will be zero: $½(-\sin\theta)^2 + ½\cos^2\theta + (1/\sqrt{2})(-\sin\theta)\cos\theta\,\overset{\text{Bose enhancement factor}}{\sqrt{2}} = ½(1 - \sin 2\theta),$ (30)

$\sqrt{2}$ factors are needed here, because each 2-photon state can split between $\hat{x}'$ and $\hat{y}'$ in two different ways

one: $½\sqrt{2}\cos\theta\,(-\sin\theta) + ½\sqrt{2}\cos\theta\sin\theta + (1/\sqrt{2})(\cos^2\theta - \sin^2\theta) = \cos 2\theta/\sqrt{2},$ (31)

two: $½\cos^2\theta + ½\sin^2\theta + (1/\sqrt{2})\cos\theta\sin\theta\,\sqrt{2} = ½(1 + \sin 2\theta).$ (32)

Thus, the emerging state from port 4 is    Bose enhancement factor

$$|\psi\rangle = ½(1 - \sin 2\theta)|0\rangle_{x'} + (1/\sqrt{2})\cos 2\theta\,|1\rangle_{x'} + ½(1 + \sin 2\theta)|2\rangle_{x'}. \tag{33}$$

The expected number of photons in port 4 is now seen to be

$$\langle n \rangle = 2[½(1 + \sin 2\theta)]^2 + [(1/\sqrt{2})\cos 2\theta]^2 = 1 + \sin 2\theta. \tag{34}$$

This is the same as $\langle n \rangle = 2\cos^2(45° - \theta)$ that one would expect based on a classical analysis.

---

**Example 4.** Let the 3-photon state $|3\rangle_x$ arrive at a $\lambda/2$-plate whose fast and slow axes, $\hat{x}'$ and $\hat{y}'$, are rotated by 45° around the $z$-axis. The amplitude that all three photons experience the plate's fast axis is $(1/\sqrt{2})^3$, in which case they emerge with their polarization along the plate's $\hat{x}'$ axis. There is a similar $(-1/\sqrt{2})^3$ amplitude that the three photons will experience the slow axis, emerging with a phase $e^{i3\pi} = -1$ 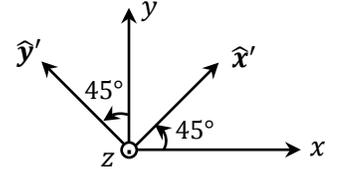
and polarization aligned with the $\hat{y}'$ axis. Both events will thus occur with a probability amplitude of $2^{-3/2}$. The third possibility is for two photons to experience the fast axis and the remaining one the slow axis, in which case the slow axis imparts a $\pi$-phase to the emergent state ($e^{i\pi} = -1$). The amplitude of this event is $(1/\sqrt{2})^2(-1/\sqrt{2})e^{i\pi}\sqrt{3} = \sqrt{3/8}$, which includes the $\sqrt{3}$ Bose enhancement factor. The enhancement is rooted in the fact that each photon has a ⅓ probability to be the one that experiences the slow axis; the corresponding amplitude is $1/\sqrt{3}$ and, since the photons are indistinguishable, these three amplitudes must be added together, thus yielding an overall enhancement factor of $3/\sqrt{3} = \sqrt{3}$. The fourth and final possibility is for two photons to experience the plate's slow axis and the remaining one the fast axis, again with an amplitude of $\sqrt{3/8}$. All in all, the emergent state will be

$$|\psi\rangle = 2^{-3/2}|3\rangle_{x'}|0\rangle_{y'} + 2^{-3/2}|0\rangle_{x'}|3\rangle_{y'} + \sqrt{3/8}\,|2\rangle_{x'}|1\rangle_{y'} + \sqrt{3/8}\,|1\rangle_{x'}|2\rangle_{y'}. \tag{35}$$

This 3-photon state is seen to be the normalized version of $[(\hat{a}^\dagger_{x'} + \hat{a}^\dagger_{y'})/\sqrt{2}\,]^3|0\rangle_{x'}|0\rangle_{y'}$, which is what one obtains by creating three photons with a linear polarization rotated 45° away from the $x'$-axis. Thus, the emergent beam is linearly-polarized along the original $y$-axis, confirming that the $\lambda/2$-plate does indeed rotate the incident photon-triplet's polarization by 90° around the $z$-axis.

---

**Example 5)** Shown in Fig.3 is a variant of the Mach-Zehnder interferometer[7,11,13] consisting of a polarizing beam-splitter (PBS) at the entrance port, a half-wave plate in the device's upper arm, a retro-reflector in the lower arm, and a 50/50 regular beam-splitter (RBS) at the exit port. The incoming wavepacket is in a 2-photon number-state consisting of one RCP and one LCP photon; see Eq.(26). The PBS directs the $x$-polarized photons to the lower arm, and the $y$-polarized photons to the upper arm of the device, creating an equal superposition of 2-photon wavepackets in the two arms. The half-wave plate in the upper arm rotates the $y$-polarized photon pair by 90°, sending the resulting $x$-polarized packet in the number-state $|2\rangle_x$ to the input port 1 of the RBS. In the lower arm, the retro-reflector is finely adjusted to eliminate any optical-path-length difference between the two arms of the interferometer. Thus, the wavepacket in the lower arm, also in the number-state $|2\rangle_x$, arrives at the entrance port 2 of the RBS, simultaneously and with zero phase-shift relative to the packet that has reached entrance port 1.

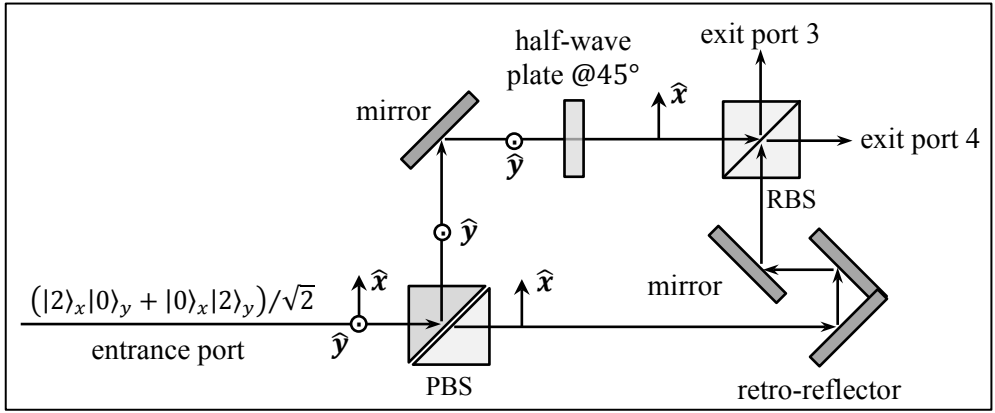

**Fig.3**. A wavepacket in the 2-photon state $(|2\rangle_x|0\rangle_y + |0\rangle_x|2\rangle_y)/\sqrt{2}$ arrives at a polarizing beam-splitter that transmits the $x$-polarized component of the incident light while reflecting the $y$-polarized component. The half-wave plate in the upper arm of the interferometer rotates the incoming polarization by 90°, rendering the pair of wavepackets that enter the 50/50 regular beam-splitter (located in the upper right-hand corner of the device) similarly polarized. When the optical-path-length difference between the two arms of the interferometer is set to zero, the 2-photon state breaks up into a pair of single-photon states that emerge from the exit ports of the regular beam-splitter.

Let the reflection and transmission coefficients of the RBS be $\rho = 1/\sqrt{2}$ and $\tau = i/\sqrt{2}$, respectively. The probability amplitude that two photons emerge from one or the other exit port of the splitter is $(\rho^2 + \tau^2)/\sqrt{2} = 0$. The only possible outcome, therefore, is for a single photon to emerge out of each exit port of the RBS. This assertion can also be proven formally, if we denote by $\hat{a}_1, \hat{a}_2, \hat{a}_3, \hat{a}_4$ the annihilation operators at the entrance and exit ports of the RBS, and recall that $\hat{a}_1^\dagger = \rho\hat{a}_3^\dagger + \tau\hat{a}_4^\dagger$ and $\hat{a}_2^\dagger = \tau\hat{a}_3^\dagger + \rho\hat{a}_4^\dagger$. Given the input state of the RBS as ½$(\hat{a}_1^\dagger\hat{a}_1^\dagger + \hat{a}_2^\dagger\hat{a}_2^\dagger)|0\rangle|0\rangle$, the corresponding output state will be

$$½\big[(\rho\hat{a}_3^\dagger + \tau\hat{a}_4^\dagger)^2 + (\tau\hat{a}_3^\dagger + \rho\hat{a}_4^\dagger)^2\big]|0\rangle|0\rangle = 2\rho\tau\hat{a}_3^\dagger\hat{a}_4^\dagger|0\rangle|0\rangle = i|1\rangle_3|1\rangle_4. \tag{36}$$

In this way, the 2-photon state entering the interferometer as depicted in Fig.3, splits into a pair of single-photon states (both linearly polarized) at the exit ports of the device.

**7. Single-photon focusing by a paraboloidal mirror**. Figure 4 shows a paraboloidal mirror whose vertex is at the origin of the $xyz$ coordinate system, whose focal point $F$ is at $(x, y, z) = (0, 0, f)$, and whose clear aperture is a circle of radius $R$ parallel to the $xy$-plane. The profile of the mirror is specified by the function $z(x, y) = \alpha(x^2 + y^2)$, where the constant coefficient $\alpha$ is related to the mirror's focal length via $\alpha = (4f)^{-1}$. Let a point source located at $\boldsymbol{r}_1 = (x_1, y_1, z_1)$, not too far from the optical axis $z$, emit a single photon of frequency $\omega$ (vacuum wavelength $\lambda = 2\pi c/\omega$). We assume that the photon can hit any point $\boldsymbol{r} = (x, y, z)$ on the mirror surface—with equal likelihood for all points $(x, y)$ within a circle of radius $R$ — then bounce back and reach some point $\mathbf{r}_2 = (x_2, y_2, z_2)$, again not too far away from the z-axis. Denoting the distance from $\mathbf{r}_1$ to $\mathbf{r}$ by $L_1$, and that from $\mathbf{r}$ to $\mathbf{r}_2$ by $L_2$, the probability amplitude for this particular path taken by the photon is given by $\exp[\mathrm{i}2\pi(L_1 + L_2)/\lambda].^3$

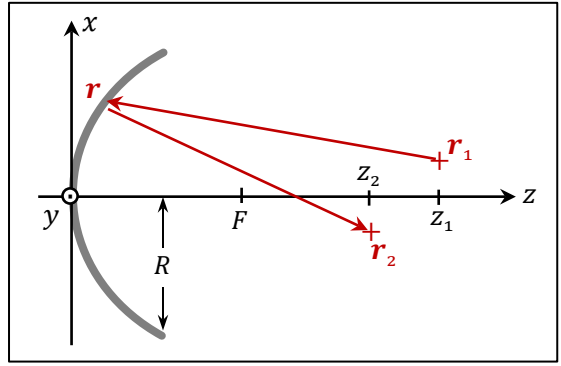

**Fig.4**. A single photon emanating from a point source located at $\mathbf{r}_1 = (x_1, y_1, z_1)$ bounces off a paraboloidal mirror at $\mathbf{r} = (x, y, z)$, then lands at $\mathbf{r}_2 = (x_2, y_2, z_2)$. The profile of the mirror is specified by the function $z(x, y) = (x^2 + y^2)/4f$, where $f$ is the mirror's focal length (i.e., the distance from the vertex at the origin of coordinates to the focal point $F$). The clear aperture of the mirror is a circle of radius $R$ and its numerical aperture is defined as $NA = R/f$. According to classical optics, if the object plane is the $xy$-plane at $z_1$, then the image plane will be the $xy$-plane located at $z_2$, where $(1/z_1) + (1/z_2) = 1/f$. Also, the geometric-optical image of the point source at $\mathbf{r}_1$ will be at $\mathbf{r}_2 = -(z_2/z_1)x_1\hat{\mathbf{x}} - (z_2/z_1)y_1\hat{\mathbf{y}} + z_2\hat{\mathbf{z}}$. In quantum optics, however, there is a non-zero probability amplitude for the photon leaving the source at $\mathbf{r}_1$ to hit the mirror at any point $\mathbf{r}$ on its paraboloidal surface, then land at an arbitrary point $\mathbf{r}_2$ in the (geometric-optical) image plane.

Let us examine the total path-length of the photon from the source at $\mathbf{r}_1$ to the mirror at $\mathbf{r}$ to the observation point at $\mathbf{r}_2$, namely,

$$L_1 + L_2 = \sqrt{(x_1 - x)^2 + (y_1 - y)^2 + (z_1 - z)^2} + \sqrt{(x_2 - x)^2 + (y_2 - y)^2 + (z_2 - z)^2}. \quad (37)$$

Each square root appearing in Eq.(37) may be approximated using $\sqrt{1 + \varepsilon} \cong 1 + \tfrac{1}{2}\varepsilon$, which is quite accurate for sufficiently small values of $\varepsilon$. Taking $z_1$ and $z_2$ to be much greater than the other parameters in Eq.(37), we write

$$L_1 = z_1\left[1 + \tfrac{(x_1^2 + y_1^2) - 2(xx_1 + yy_1 + zz_1) + (x^2 + y^2 + z^2)}{z_1^2}\right]^{1/2} \cong z_1 + \tfrac{(x_1^2 + y_1^2) - 2(xx_1 + yy_1) + (x^2 + y^2 + z^2)}{2z_1} - z. \quad (38)$$

A similar approximation applies to $L_2$ as well. We thus have

$$L_1 + L_2 \cong z_1 + z_2 + \tfrac{x_1^2 + y_1^2}{2z_1} + \tfrac{x_2^2 + y_2^2}{2z_2} - \left(\tfrac{x_1}{z_1} + \tfrac{x_2}{z_2}\right)x - \left(\tfrac{y_1}{z_1} + \tfrac{y_2}{z_2}\right)y + \tfrac{1}{2}\left(\tfrac{1}{z_1} + \tfrac{1}{z_2}\right)(x^2 + y^2 + z^2) - 2z. \quad (39)$$

The first four terms of the total path-length in Eq.(39) are independent of $x$ and $y$. Therefore, they only contribute a constant phase to the overall probability amplitude at a given $\mathbf{r}_2$, namely,

$$e^{\mathrm{i}\varphi_0} = \exp\left[\mathrm{i}2\pi\left(\tfrac{z_1 + z_2}{\lambda} + \tfrac{x_1^2 + y_1^2}{2\lambda z_1} + \tfrac{x_2^2 + y_2^2}{2\lambda z_2}\right)\right]. \quad (40)$$

With regard to the point $\mathbf{r}$ on the surface of the paraboloidal mirror, we have $z = \alpha(x^2 + y^2)$, which, upon substitution into Eq.(39), further simplifies the remaining terms of $L_1 + L_2$, as follows:

$$\tfrac{1}{2}\left(\tfrac{1}{z_1} + \tfrac{1}{z_2} - 4\alpha\right)(x^2 + y^2) + \tfrac{1}{2}\left(\tfrac{1}{z_1} + \tfrac{1}{z_2}\right)[\alpha(x^2 + y^2)]^2 - \left[\left(\tfrac{x_1}{z_1} + \tfrac{x_2}{z_2}\right)x + \left(\tfrac{y_1}{z_1} + \tfrac{y_2}{z_2}\right)y\right]. \quad (41)$$

The first term of this expression now reveals that the focal length of the paraboloidal mirror must be $f = (4\alpha)^{-1}$, since, by requiring $z_1$ and $z_2$ to satisfy the identity $(1/z_1) + (1/z_2) = 1/f$, that term disappears. The remaining terms then become

$$\frac{(x^2+y^2)^2}{32f^3} - \left[\left(\frac{x_1}{z_1} + \frac{x_2}{z_2}\right)x + \left(\frac{y_1}{z_1} + \frac{y_2}{z_2}\right)y\right]. \tag{42}$$

The leading term of the above expression corresponds to the radial phase profile $\varphi(\rho) = 2\pi\rho^4/(32\lambda f^3)$, with $\rho = \sqrt{x^2 + y^2}$, which represents a form of spherical aberration.[10,11] Recalling that the numerical aperture $NA$ of the mirror is the ratio of its aperture radius $R$ to its focal length $f$, the maximum value of $\varphi(\rho)$ occurs at $\rho = R$ and is given by $\varphi(R) = (\pi/16)(f/\lambda)NA^4$. Thus, for reasonably small values of $NA$ and $f/\lambda$ (say, $NA = 0.05$, $f = 20\ cm$, $\lambda = 0.5\ \mu m$), the first term of the expression (42) may be ignored. As for the remaining terms, given that the geometric-optical image of the point source at $(x_1, y_1, z_1)$ appears at $(x_2, y_2, z_2)$, where $(x_2, y_2) = -(z_2/z_1)(x_1, y_1)$, we redefine the $xy$ coordinates at $z = z_2$ relative to the geometric image of the point source; that is,

$$\tilde{x}_2 = x_2 - (z_2/z_1)x_1 \quad \text{and} \quad \tilde{y}_2 = y_2 - (z_2/z_1)y_1. \tag{43}$$

Thus, the total probability amplitude that the photon emitted at $(x_1, y_1, z_1)$ will bounce off the mirror at an unknown location, then proceed to land at $(\tilde{x}_2, \tilde{y}_2, z_2)$ is found to be

$$\text{Amplitude at } (\tilde{x}_2, \tilde{y}_2, z_2) = \iint_{\substack{\text{circle of}\\\text{radius } R}} e^{-i2\pi(\tilde{x}_2 x + \tilde{y}_2 y)/(\lambda z_2)} \mathrm{d}x\mathrm{d}y$$

$$= \int_{\rho=0}^{R}\int_{\theta=0}^{2\pi} e^{-i2\pi\tilde{\rho}_2\rho\cos(\theta-\tilde{\theta}_2)/\lambda z_2} \rho\mathrm{d}\theta\mathrm{d}\rho \quad \leftarrow \boxed{\tilde{\rho}_2 = \sqrt{\tilde{x}_2^2 + \tilde{y}_2^2}}$$

$$\boxed{\int_0^{2\pi} e^{\pm ix\cos\theta}\mathrm{d}\theta = 2\pi J_0(x)} \rightarrow = 2\pi\int_{\rho=0}^{R}\rho J_0(2\pi\tilde{\rho}_2\rho/\lambda z_2)\mathrm{d}\rho = \frac{RJ_1(2\pi R\tilde{\rho}_2/\lambda z_2)}{\tilde{\rho}_2/\lambda z_2}. \leftarrow \boxed{\int xJ_0(x)\mathrm{d}x = xJ_1(x)} \tag{44}$$

The probability amplitude of the single photon arriving in the image plane at $(\tilde{x}_2, \tilde{y}_2, z_2)$ is thus seen to coincide with the well-known Airy pattern of classical physical optics.[10,11]

**8. Concluding remarks**. A pair of coherent (i.e., quasi-classical or Glauber) beams $|\gamma_1\rangle$ and $|\gamma_2\rangle$, entering through ports 1 and 2 of the lossless, regular beam-splitter depicted in Fig.1, can be shown to emerge as coherent beams $|\rho\gamma_1 + \tau\gamma_2\rangle$ and $|\tau\gamma_1 + \rho\gamma_2\rangle$ from the exit ports 3 and 4, respectively. This can be proven by invoking the creation operator $e^{-|\gamma|^2/2}e^{\gamma\hat{a}^\dagger}$ for a coherent beam having an arbitrary (complex) parameter $\gamma$ in conjunction with the operator identities given by Eq.(10). An alternative proof that does not rely on the operator algebra requires nothing more than summing up the probability amplitudes of the various emergent number-states at the exit ports corresponding to a superposition of all the number-states $|n_1\rangle|n_2\rangle$ arriving at the entrance ports of the RBS. Using similar methods, one can demonstrate that combinations of coherent beams in different polarization states continue to be coherent. Consider, for instance, a pair of co-propagating coherent beams $|\gamma_1\rangle_x$ and $|\gamma_2\rangle_y$, one linearly-polarized along $\hat{x}$, the other along $\hat{y}$. Given that the creation operators $\hat{a}_x^\dagger$ and $\hat{a}_y^\dagger$ for $x$- and $y$-polarized photons commute, the creation operators for the pair of linearly-polarized coherent beams combine to yield the overall creation operator, as follows:

$$e^{-(|\gamma_1|^2+|\gamma_2|^2)/2}\exp(\gamma_1\hat{a}_x^\dagger + \gamma_2\hat{a}_y^\dagger). \tag{45}$$

The operator appearing in the exponent may now be written as

$$\gamma_1\hat{a}_x^\dagger + \gamma_2\hat{a}_y^\dagger = \sqrt{|\gamma_1|^2 + |\gamma_2|^2}e^{i\varphi_{\gamma_1}}\left[\cos\theta\ \hat{a}_x^\dagger + e^{i(\varphi_{\gamma_2}-\varphi_{\gamma_1})}\sin\theta\ \hat{a}_y^\dagger\right], \tag{46}$$

where $\cos\theta = |\gamma_1|/\sqrt{|\gamma_1|^2 + |\gamma_2|^2}$ and $\sin\theta = |\gamma_2|/\sqrt{|\gamma_1|^2 + |\gamma_2|^2}$. The combined creation operator is seen to be that of a coherent state, with amplitude parameter $\gamma = \sqrt{|\gamma_1|^2 + |\gamma_2|^2}\,e^{i\varphi_{\gamma_1}}$ and elliptical polarization specified by the angle $\theta$ and the relative phase $\varphi_{\gamma_2} - \varphi_{\gamma_1}$ between $\hat{x}$ and $\hat{y}$ components.

In our discussion of Sec. 7, we pointed out the existence of a small amount of spherical aberration on the probability amplitude of the reflected photon at the mirror surface — which we subsequently proceeded to ignore. A certain amount of aberration is, of course, expected to accompany a classical wavefront reflected from a paraboloidal mirror when the object is off-axis or located at a finite distance from the mirror. However, the spherical aberration encountered in Sec. 7 does *not* diminish when the point source is far from the mirror and centered on the optical axis. The unexpected appearance of this aberration (negligible though it may be under the stated circumstances) is an inevitable consequence of the paraxial approximation introduced early in Sec. 7.

The goal of the present paper has been to showcase several examples from the field of quantum optics that are amenable to less formalistic and more intuitive analyses based on direct applications of the Feynman principle. Additional examples can be found in some of our recent publications.[15-18]

**Appendix**

It is well known that the Fresnel reflection and transmission coefficients ($\rho$ and $\tau$) of a lossless regular beam-splitter (RBS) are complex numbers that satisfy the relations $|\rho|^2 + |\tau|^2 = 1$ and $\varphi_\rho = \varphi_\tau \pm 90°$. Proofs of these properties based on the principle of conservation of energy and the time-reversal symmetry of Maxwell's equations can be readily found in the literature.[7-9,13] Here, we present an elementary demonstration of the 90° phase difference between $\rho$ and $\tau$ in the special case where the beam-splitter is a large, thin, uniform sheet, made of a non-absorbing dielectric material.

Figure A1(a) shows a linearly-polarized plane-wave arriving at oblique incidence on the thin sheet. The incident wave's electric field is $\boldsymbol{E}_0 e^{i(\boldsymbol{k}\cdot\boldsymbol{r} - \omega t)}$, where $\boldsymbol{E}_0 = E_{0y}\hat{\boldsymbol{y}}$ is the field amplitude, $\boldsymbol{k} = k_x\hat{\boldsymbol{x}} + k_z\hat{\boldsymbol{z}}$ is the $k$-vector, and $\omega$ is the frequency of the monochromatic oscillations; this is the case of an *s*-polarized incident light.[12] The excited electric dipoles within the sheet radiate into the surrounding space a pair of identical plane-waves with $E$-fields given by $\rho E_{0y}\hat{y}e^{i(k_x x \pm k_z z - \omega t)}$. Here, the plus and minus signs correspond to the emergent beams on the right- and left-hand sides of the sheet, respectively, and the coefficient $\rho$ is the sheet's Fresnel reflection coefficient.[14] On the transmission side of the sheet, the undiminished incident plane-wave continues its journey in parallel with (and super-imposed onto) the field that is being radiated by the excited dipoles into the half-space on the right-hand side, which means that the Fresnel transmission coefficient of the sheet must be $\tau = 1 + \rho$.

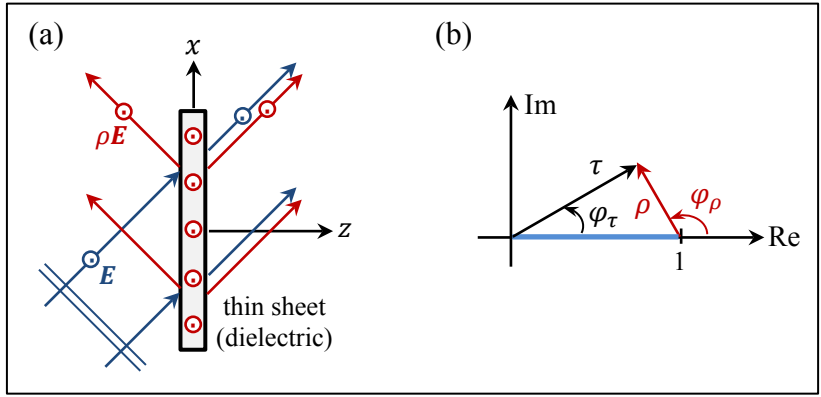

**Fig. A1**. (a) A linearly-polarized plane-wave shines obliquely on a thin sheet of a homogeneous, isotropic, linear, non-absorbing material. The excitation of atomic dipoles within the sheet gives rise to a pair of identical plane-waves that propagate away from the sheet in opposite directions. The plane-wave propagating backwards (i.e., toward the source of the incident radiation) constitutes the reflected beam whose $E$-field amplitude is $\rho$ times that of the incident wave. In the forward (i.e., transmission) direction, the radiated beam combines with the (undiminished) continuation of the incident beam to produce the transmitted plane-wave whose $E$-field amplitude is $1 + \rho$ times that of the incident wave. (b) In the complex plane, the transmission coefficient $\tau = 1 + \rho$ forms one side of a right-angle triangle whose other side is $\rho$ and whose hypotenuse has length 1, as demanded by the requirement of energy conservation, namely, $|\rho|^2 + |\tau|^2 = 1$. The phase angles of $\rho$ and $\tau$ are seen to be related by $\varphi_\rho = \varphi_\tau + 90°$.

Figure A1(b) is a complex-plane diagram of the Fresnel coefficients $\rho$ and $\tau$. Conservation of energy demands that $|\rho|^2 + |\tau|^2 = 1$, indicating that the triangle formed by $\rho$, $\tau$, and the unit-length segment of the real axis is a right-angle triangle. The phase angles of $\rho$ and $\tau$ are thus seen to satisfy the identity $\varphi_\rho = \varphi_\tau + 90°$. (If $\varphi_\rho$ and $\varphi_\tau$ happen to be negative, the relation will be $\varphi_\rho = \varphi_\tau - 90°$.)

**Acknowledgement**. The author is grateful to Ewan Wright, Brian Anderson, Dalziel Wilson, and José Sasián of the Wyant College of Optical Sciences, University of Arizona, and also to Tobias Mansuripur of *Pendar Technologies*, for numerous fruitful discussions.